\documentclass[12pt, reqno]{amsart}

\usepackage{amssymb} 

\newcommand{\ii}{\mathrm{i}}
\newcommand{\wt}{\widetilde}
\newcommand{\wh}[1]{\widehat{#1}}
\newcommand{\ov}[1]{\overline{#1}}

\newcommand{\myqed}{\hfill $\square$}
\newcommand{\mydiamond}{\hfill $\Diamond$}

\newtheorem{prop}{Proposition}

\begin{document} 

\title[Fluctuations of intensive quantities]{Fluctuations of intensive quantities in statistical thermodynamics}
\author[A. E. Ruuge]{Artur E. Ruuge}
\address{
Department of Mathematics and Computer Science, 
University of Antwerp, 
Middelheim Campus Building G, 
Middelheimlaan 1, B-2020, 
Antwerp, Belgium
}
\email{artur.ruuge@uantwerpen.be}

\begin{abstract}
In phenomenological thermodynamics,  
the canonical coordinates of a physical system split in pairs with  
each pair consisting of 
an extensive quantity and an intensive one. 
In the present paper 
the quasithermodynamic fluctuation theory   
of a model system of a large number of oscillators 
is extended 
to statistical thermodynamics based on the idea to perceive 
the fluctuations of intensive variables as the fluctuations of specific extensive ones 
in a ``thermodynamically dual'' system.   
The extension is motivated by the symmetry of the problem in the context of an 
analogy with quantum mechanics which is 
stated in terms of a generalized Pauli problem for the thermodynamic fluctuations. 
The doubled Boltzmann constant divided by the number of particles plays a similar role to 
the Planck constant.

\end{abstract}

\maketitle

\section{Introduction}

In quantum mechanics, it is a standard practice 
to perceive the Planck constant, $\hbar$, as a small parameter 
of the semiclassical approximation \cite{MaslovFedoriuk, Leray}
and to write $\hbar \to 0$. 
One should keep in mind that, in principle, 
the symbol, $\hbar$, denotes a fundamental \emph{constant}, which
has a nontrivial physical dimension: 
\begin{equation} 
\label{eq:Planck_constant} 
\hbar = 1.054 5716 \times 10^{- 27} \mathrm{erg} \cdot \mathrm{s}
\end{equation} 

The smallness of $\hbar$ 
must be perceived in comparison to the 
``classical action'', {\it i.e.}, to the typical values of 
the classical action variables in the Hamilton-Jacobi formalism. 

In statistical thermodynamics, the Boltzmann constant:
\begin{equation} 
\label{eq:Boltzmann_constant}
k_{B} = 1.380 6488 \times 10^{- 16} \mathrm{erg} \cdot \mathrm{K}^{-1}
\end{equation} 
is a natural candidate for a similar role. 
The notation, $k_{B} \to 0$, looks just like $\hbar \to 0$, 
but it is relatively uncommon. 
Nonetheless, it can 
be defined in such a way that 
the corresponding limit transition is equivalent to $N \to \infty$, 
the large number of particles limit 
restricted by the condition that 
the values of specific extensive quantities 
(e.g., average energy per particle) remain fixed. 
One may compare this to an equivalent point of view on $\hbar \to 0$, 
which is also termed 
the ``large quantum numbers'' limit in quantum mechanics. 
\vspace{6pt}

It is convenient to separate 
three levels of description of an abstract mechanical system: 
\begin{itemize} 
\item[$L1$:] Classical mechanics; 

\item[$L2$:] Semiclassical mechanics; 

\item[$L3$:] Quantum mechanics. 
\end{itemize}

In thermodynamics, there are also three levels: 
\begin{itemize}
\item[$\Lambda 1$:] Phenomenological thermodynamics; 

\item[$\Lambda 2$:] Quasithermodynamics; 

\item[$\Lambda 3$:] Statistical thermodynamics. 
\end{itemize}

Quasithermodynamics is the theory of fluctuations of thermodynamic quantities if 
the number of particles in the system, $N \gg 1$, is large, but not ``huge'', 
and statistical thermodynamics is the microscopic theory of \emph{heat} 
associated with such concepts as the Gibbs distribution and the constant, $k_B$. 
Phenomenological thermodynamics is described by the facts that 
the amount of a chemical substance in a system is measured in moles $\nu$ and 
not as a number of particles, $N$. 
The Boltzmann constant is not introduced into the theory yet, 
but the universal gas constant:
\begin{equation}
\label{eq:universal_gas_constant}
R = 8.314 4621 \times 10^{7} \, 
\mathrm{erg} \cdot \mathrm{K}^{-1} \cdot \mathrm{mol}^{-1}
\end{equation}
is well defined. 

It is a rather special coincidence that 
two so different theories, like classical mechanics and phenomenological thermodynamics, 
are closely related to such a fundamental mathematical concept, like the \emph{Lagran\-gian manifold}. 
Moreover, a further analysis 
\cite{Maslov_thermo1, Maslov_thermo2, Maslov_thermo3}
leads to an idea 
of axiomatizing the asymptotic expansions of the partition function 
in terms of the \emph{tunnel canonical operator} 
\cite{Maslov_asymp_meth, Maslov_Nazaikinskii}. 

Take an abstract phenomenological thermodynamic system with extensive coordinates 
$E = (E_0, E_1, \dots, E_d)$. The entropy function 
$S = S (E)$ satisfies 
$S (\lambda E_0, \lambda E_1, \dots, \lambda E_d) = \lambda S (E)$, 
for all $\lambda > 0$. 
According to the first law of thermodynamics: 
\begin{equation} 
d S (E) = \beta_0 d E_0 + \beta_1 d E_1 + \dots + \beta_d d E_d
\end{equation} 
where $\beta = (\beta_0, \beta_1, \dots, \beta_d)$ is the 
collection of the corresponding intensive coordinates. 
For example, 
take a one component system with $\nu$ moles of chemical substance 
described by the coordinates, $E_0 = \nu$, the internal energy, $E_1$, and the volume, $E_2$, $d = 2$. 
Then, $\beta_0 = - \mu/ T$, $\beta_1 = 1/ T$, and $\beta_2 = p/ T$, 
where $\mu$ is chemical potential, 
$T$ is absolute temperature and 
$p$ is pressure. 

Assume that $S = S (E)$ is a smooth function over a domain, 
$\mathcal{D} \subset \mathbb{R}^{d + 1} (E)$, and 
consider a manifold, $\Lambda_{S} \subset \mathbb{R}^{2 (d + 1)} (\beta, E)$: 
\begin{equation} 
\Lambda_{S} = \lbrace (\beta, E) 
\,|\, \beta_j = \partial S (E)/ \partial E_j, \, E \in \mathcal{D}, j = 0, 1, \dots, d \rbrace 
\end{equation}
Let $i: \Lambda_S \to \mathbb{R}^{2 (d + 1)} (\beta, E)$ be the canonical embedding, and 
put:
\begin{equation}
\omega = \sum_{j = 0}^{d} d \beta_j \wedge d E_j
\end{equation}

The two-form $\omega$ defines a symplectic structure on $\mathbb{R}^{2 (d + 1)} (\beta, E)$, and 
we have: $i^{*} (\omega) = 0$, {\it i.e.}, $\Lambda_S$ is a Lagrangian manifold with respect to $\omega$. 
Note that $\omega$ does not depend on $S$, and therefore, $\Lambda_S$ is a Lagrangian manifold for any $S$. 

The Lagrangian manifold, $\Lambda_{S}$ in thermodynamics is \emph{connected} and \emph{simply connected}; 
plus, there is a condition at infinity corresponding to the third law of thermodynamics. 
Furthermore, $\Lambda_{S}$ is covered by a single global chart with coordinates $E = (E_0, E_1, \dots, E_d)$. 
The states of the thermodynamic equilibrium of the system are in one-to-one correspondence with the points of $\Lambda_S$, 
and it is convenient to lift $S = S (E)$ to a function $\wh{S} = \wh{S} (\alpha)$ on 
the Lagrangian manifold, $\alpha \in \Lambda_{S}$, so that 
$d \wh{S} = i^{*} (\sum_{j = 0}^{d} \beta_j d E_j)$. 
If we perceive $E_j$, $j = 0, 1, \dots, d$ 
as an analogue of generalized coordinates in semiclassical mechanics, then 
the entropy is an \emph{action} on the 
Lagrangian manifold of the equilibrium states of a thermodynamic system in the $E$-chart.

Consider now an abstract mechanical system with $D$ degrees of freedom described by 
coordinates $q = (q_1, q_2, \dots, q_{D})$ and the canonically conjugate 
momenta $p = (p_1, p_2, \dots, p_{D})$ 
corresponding to the symplectic structure 
$\omega_{\mathrm{cl}} = \sum_{i = 1}^{D} d p_i \wedge d q_i$ on 
the phase space $\mathbb{R}^{2 D} (p, q)$. 
Take a Lagrangian manifold, $\Lambda_{\mathrm{cl}} \subset \mathbb{R}^{2 D} (p, q)$, 
and assume that $U \subset \Lambda_{\mathrm{cl}}$ is a $q$-chart on $\Lambda_{\mathrm{cl}}$. 
In \emph{semiclassical} mechanics, 
the action $S_{\mathrm{cl}} = S_{\mathrm{cl}} (q)$ in $U$
has another important interpretation: 
it corresponds to the phase of the fast oscillating exponent 
in the 
Wentzel-Kramers-Brillouin (WKB) ansatz 
for the wavefunction, $\psi_{\hbar} (q)$, in the Schr\"odinger equation, 
$\psi_{\hbar} (q) \sim \exp (\ii S_{\mathrm{cl}} (q)/ \hbar)$. 
It is natural to expect that the semiclassical $\hbar \to 0$ methods of 
quantum mechanics 
can be ``transplanted'' to a thermodynamic Lagrangian manifold, $\Lambda_{S}$, and that 
this might lead to some new insights about statistical thermodynamics.

There exists a certain similarity between the quasithermodynamic fluctuation theory and the 
Heisenberg uncertainty relation, which has attracted the attention of many researchers 
\cite{Velazquez, Kazinski, Lavenda, RudoiSukhanov, UffinkLith, VelazquezCurilef, Mandelbrot, Ruppeiner}. 
This fact has already been known to N.~Bohr and W.~Heisenberg, 
who have tried to extend the philosophical concept of \emph{complementarity} to thermodynamics. 
The corresponding analogy is far from being straightforward, since it 
depends strongly on the interpretation of the nature of intensive thermodynamic quantities. 
In \cite{Ruuge1, Ruuge2, Ruuge3},
it is suggested to ``span'' a Pauli problem 
\cite{Ibort_Manko_Marmoc_Simonic_Ventrigliac, Mancini_Manko_Tombesi} 
over the fluctuations of thermodynamic quantities 
considered in the quasithermodynamic approximation. 
The similarities in the mathematical formalism 
of the classical mechanics and phenomenological thermodynamics 
are quite of interest from the perspective of relativistic quantum physics 
\cite{Balian_Valentin, Rovelli, Connes_Rovelli, Montesinos_Rovelli, Rajeev_contact}, 
as well as in ``quantum thermodynamics'' in the sense of 
\cite{Henrich_Michel_Mahler, Quan_Liu_Sun_Nori, Linden_Popescu_Short_Winter, Skrzypczyk_Brunner_Linden_Popescu}. 
There are also 
attempts to extend the analogy with quantum mechanics to 
non-equilibrium thermodynamics 
\cite{Onsager_Machlup, Machlup_Onsager, Mehrafarin, Acosta_Fernandez-de-Cordoba_Isidro_Santander, Fernandez-de-Cordoba_Isidro_Perea, Ruuge4}. 
It is worth mentioning that numerical simulations 
of non-equilibrium multiparticle systems 
on a computer admit a natural parallelization \cite{InozemtsevaPerepelkinSadovnikov}. 

The present work focuses on the \emph{equilibrium} case. 
In quasithermodynamics, the intensive quantities and the \emph{specific} extensive 
quantities enter the theory in a very symmetric way. 
We consider a problem of extension of this symmetry to \emph{statistical} thermodynamics. 
The basic idea is to perceive the fluctuations of intensive quantities 
as fluctuations of specific extensive quantities in \emph{another} auxiliary 
system. 
We provide a model example of thermodynamic duality and then state a generalized Pauli problem. 
It turns out that $2 k_{B}/ N$, where $N$ is the number of particles in the model system, 
plays the same role as $\hbar$ in quantum mechanics.

\section{Quasithermodynamics and Quantization}

The Planck constant, $\hbar$, is similar to the Boltzmann constant, $k_B$, but 
there is an essential difference. 
In semiclassical mechanics, it enters 
the Bohr-Sommerfeld quantization condition, which 
for an abstract $D$-dimensional mechanical system on a 
Lagrangian torus, $\Lambda_{\mathrm{cl}} \subset \mathbb{R}^{2 D} (p, q)$, 
modulo the correction given by the Maslov index, 
is of the shape: 
\begin{equation} 
\frac{1}{2 \pi} \oint_{\gamma_k} \sum_{j = 1}^{D} p_j d q_j \sim \hbar n_{k}
\end{equation}
where $n_{k} \in \mathbb{Z}_{\geqslant 0}$, $k = 1, 2, \dots, D$ and 
$\gamma_k$ correspond to the generators of 
the fundamental group, $\pi_1 (\Lambda_{\mathrm{cl}}) \approx \mathbb{Z}^{D}$. 
For example, for a one-dimensional harmonic oscillator of frequency $\omega$ 
considered near a classical value of energy, $E$, we have: $E \omega^{-1} \sim \hbar n$, $n \in \mathbb{Z}$, $n \gg 1$. 
Intuitively, quantization is a splitting of the classical quantity, $E \omega^{-1}$, 
into a product of a ``very small'' quantity, $\hbar$, and a ``very large'' quantity,~$n$.

The Lagrangian manifold, $\Lambda_{\mathrm{cl}}$, in the Bohr-Sommerfeld formula 
is \emph{not} simply connected, 
while 
the Lagrangian manifold in thermodynamics \emph{is} simply connected. 
This fact follows directly from 
the axiomatics of phenomenological thermodynamics. 
For an abstract thermodynamic system, 
$\Lambda^{0} \subset \mathbb{R}^{2 (d + 1)} (\wt{\beta}, E)$, 
of dimension $d$, 
where $E = (E_0, E_1, \dots, E_{d})$ are the extensive coordinates and 
$\wt{\beta} = (\beta_{0}, \beta_{1}, \dots, \beta_{d})$ are the intensive coordinates, 
we have: 
\begin{equation} 
\oint_{\gamma} \sum_{j = 0}^{d} \beta_j d E_j = 0
\end{equation}
for any closed path $\gamma \subset \Lambda^{0}$. 
Therefore, the formula $\nu R \sim k_{B} N$, 
where $\nu$ is the number of moles and $N$ is the number of particles of a chemical substance 
in the system, is not exactly similar to $E \omega^{-1} \sim \hbar n$.

We should also point out that, 
since the entropy $S = S (E)$ satisfies 
$S (\lambda E_0, \lambda E_1, \dots, \lambda E_{d}) = 
\lambda S (E)$, 
for all $\lambda > 0$, 
the Lagrangian manifold $\Lambda^{0}$ has an additional property: 
if $\alpha^{0} \in \Lambda^{0}$ has coordinates $E_{j} (\alpha^0)$, $\beta_{l} (\alpha^0)$, 
$j, l = 0, 1, \dots, d$, 
then for every $\lambda > 0$, the manifold, $\Lambda^{0}$, also contains 
$\alpha_{\lambda}^{0} \in \Lambda^{0}$ with coordinates
$E_{j} (\alpha_{\lambda}^0) = \lambda E_{j} (\alpha^0)$, $\beta_{l} (\alpha_{\lambda}^0) = \beta_{l} (\alpha^0)$, 
$j, l = 0, 1, \dots, d$.

We may assume that $E_0 > 0$ on $\Lambda^{0}$. Then, it is 
convenient to introduce the \emph{specific} 
extensive coordinates 
$\varepsilon_j = E_{j}/ E_{0}$, $j = 1, 2, \dots, d$ 
and the \emph{specific} entropy $s = S/ E_0$, $s = s (\varepsilon)$, 
$\varepsilon = (\varepsilon_1, \varepsilon_2, \dots, \varepsilon_d)$. 
The Lagrangian manifold, $\Lambda^{0}$, corresponds to 
$\Lambda \subset \mathbb{R}^{2 d} (\beta, \varepsilon)$, 
$\beta = (\beta_1, \beta_2, \dots, \beta_{d})$, 
which is a Lagrangian manifold 
with respect to the 
symplectic structure $\omega = \sum_{j = 1}^{d} d \beta_j \wedge d \varepsilon_j$, 
and 
we have: 
\begin{equation}
\label{eq:first_law_specific} 
i_{\Lambda}^{*} \Big( d s - \sum_{j = 1}^{d} \beta_j d \varepsilon_j \Big) = 0
\end{equation}
where $i_{\Lambda} : \Lambda \to \mathbb{R}^{2 d} (\beta, \varepsilon)$ 
is the canonical embedding. 

The previous formula, Equation \eqref{eq:first_law_specific}, is a formula of \emph{phenomenological} thermodynamics. 
Let us now discuss what happens in \emph{quasithermodynamics}. 
Let $d = 1$, and assume that the thermodynamic system contains $\nu$ moles of a single chemical substance. 
Let $E_0 = \nu R/ k_B$; 
$R$ is the universal gas constant Equation \eqref{eq:universal_gas_constant} and 
$k_B$ is the Boltzmann constant Equation \eqref{eq:Boltzmann_constant}. Let $E_1$ denote the internal energy. 
Then, $\beta_0 = - \mu k_B/ (R T)$, and $\beta_1 = 1/ T$, where $\mu$ is the chemical potential and 
$T$ is the absolute temperature. 
It is convenient to write just $\beta$ and $\varepsilon$ in place of $\beta_1$ and 
$\varepsilon_1 = E_1/ E_0$, respectively. 

The quantity, $\varepsilon$, is termed the \emph{specific internal energy} of the system. 
Assume that $s'' (\varepsilon) < 0$ for all values of $\varepsilon$. 
Let $\phi (\beta)$ denote the Legendre transform of $s (\varepsilon)$: 
\begin{equation} 
\phi (\beta) = (- \beta \varepsilon + s (\varepsilon))|_{\varepsilon = \varepsilon (\beta)} 
\end{equation}
where $\varepsilon = \varepsilon (\beta)$ is the solution of the equation 
$- \beta + s' (\varepsilon) = 0$ with respect to $\varepsilon$. 
In quasithermodynamics, there is a large 
parameter in the formulae 
$N \to + \infty$, which corresponds to $\nu R/ k_B \in \mathbb{R}$. 
More precisely, the scheme of a computation is as follows: 
find an asymptotic expansion with respect to 
$N \to + \infty$; 
keep the required number of terms, and then replace 
$N$ 
with a numerical value, $\nu R/ k_B$. 
Take a point, $\alpha \in \Lambda$, and 
denote its coordinates in the ambient space, $\mathbb{R}^{2} (\beta, \varepsilon)$, as 
$(\beta (\alpha), \varepsilon (\alpha))$. 
The corresponding fluctuations, $\delta \beta$ and $\delta \varepsilon$, 
around the values $\beta = \beta (\alpha)$ and $\varepsilon = \varepsilon (\alpha)$, 
are described 
by the probability densities, $f_{\delta \beta} (y; \alpha, N)$, $y \in \mathbb{R}$, 
and $f_{\delta \varepsilon} (x; \alpha, N)$, $x \in \mathbb{R}$, 
which are approximated by the normal distributions of the shape: 
\begin{gather} 
\label{eq:fluct_delta_varepsilon}
f_{\delta \varepsilon} (x; \alpha, N) \simeq
\Big( \frac{2 \pi k_B}{N \lambda} \Big)^{1/ 2}
\exp \Big\lbrace - N \frac{\lambda x^2}{2 k_B} \Big\rbrace \\ 
\label{eq:fluct_delta_beta}
f_{\delta \beta} (y; \alpha, N) \simeq
\Big( \frac{2 \pi k_B \lambda}{N} \Big)^{1/ 2}
\exp \Big\lbrace - N \frac{y^2}{2 k_B \lambda} \Big\rbrace 
\end{gather}
where $\lambda = \lambda (\alpha) = - s'' (\varepsilon (\alpha)) = 1/ \phi '' (\beta (\alpha))$, 
$\alpha \in \Lambda$, $N \in \mathbb{R}$. 

The interpretation of the fluctuations, $\delta \varepsilon$, is quite straightforward 
from the perspective of the canonical Gibbs formalism, which 
we discuss in more detail in the next section. 
On the other hand, an interpretation of the fluctuations of the inverse temperature, 
$\delta \beta$, turns out to be quite problematic. 
Moreover, there are different points of views on this subject, 
some of them described 
in \cite{UffinkLith}. 

Naively, if one accepts a ``definition'' of the inverse temperature 
as the parameter, $\beta$, in the canonical Gibbs distribution, 
then the fluctuations of $\delta \beta$ do not exist at all if 
the system is placed in a thermostat. 
At the same time, there is no problem to consider the quantities like 
the variance of energy in the system or the higher moments, so the 
fluctuations, $\delta \varepsilon$, receive a natural interpretation. 
What is then the meaning of the probability density Equation \eqref{eq:fluct_delta_beta}?

In Landau-Lifshitz \cite{LandauLifshitz}, the problem with $\delta \beta$ is essentially ``swept under the carpet''. 
They denote the fluctuation of inverse temperature as $\Delta \beta$ and 
the fluctuation of the inner energy as $\Delta E$ and 
put ``by definition'' $\Delta \beta \simeq (\partial^{2} S/ \partial E^{2})_{\nu} \Delta E$, 
where $S$ is the entropy as a function of the internal energy, $E$, and the number of 
moles, $\nu$, of the chemical substance. 
Basically, there is only one independent random variable, $\Delta E$, 
associated with a selected equilibrium state $(E, \nu)$ of the system, and 
$\Delta \beta$ is just a transformation of $\Delta E$. 
In the geometric picture involving the Lagrangian manifold, $\Lambda \subset \mathbb{R}^{2} (\beta, \varepsilon)$, 
one may then intuitively think of the fluctuations as follows: 
there is a point on $\Lambda$ 
that randomly moves a little around a 
fixed position, but 
it stays always on the manifold, {\it i.e.}, 
the value of $\beta$ is immediately adjusted to the value of $\varepsilon$. 

The Equations \eqref{eq:fluct_delta_varepsilon} and \eqref{eq:fluct_delta_beta}, imply 
that:
\begin{equation}
\label{eq:thermodynamic_uncertainty}
\langle (N^{1/2} \delta \varepsilon)^{2} \rangle_{N, \alpha} \langle (N^{1/2} \delta \beta)^2 \rangle_{N, \alpha} \simeq k_B^2
\end{equation}
where $\langle - \rangle_{N, \alpha}$ denotes taking the average with respect to 
the distributions in Equations \eqref{eq:fluct_delta_varepsilon} and \eqref{eq:fluct_delta_beta}. 
This is formally similar to the Heisenberg uncertainty relation in quantum mechanics, and 
we notice 
that $\hbar$ corresponds to the \emph{doubled} Boltzmann constant, $2 k_B$. 
At the same time, if we accept the Landau-Lifshitz point of view, then 
the linear correlation coefficient between $N^{1/ 2} \delta \beta$ and 
$N^{1/ 2} \delta \varepsilon$ 
is equal to $- 1$. 
Take an abstract one-dimensional quantum system with a coordinate $\wh{q} = x$ and the 
canonically conjugate momentum $\wh{p} = - \ii \hbar \partial/ \partial x$, in a coherent state 
with a wavefunction :
\begin{equation} 
\label{eq:coherent_state}
\psi_{\hbar} (x; p_0, q_0, \lambda) = 
\Big( \frac{2 \pi \hbar}{\lambda} \Big)^{1/ 4}
\exp (\ii p_0 x/ \hbar) \exp \Big\lbrace - \frac{\lambda (x - q_0)^2}{4 \hbar} \Big\rbrace 
\end{equation}
where $p_0, q_0 \in \mathbb{R}$ and $\lambda > 0$ are parameters. 
The corresponding quantum averages are 
$\langle \wh{q} \rangle_{\hbar} = q_0$ and 
$\langle \wh{p} \rangle_{\hbar} = p_0$, 
and for the fluctuations, 
$\delta \wh{q} = \wh{q} - q_0$ and $\delta \wh{p} = \wh{p} - p_0$;
we have: 
$\langle (\delta \wh{q})^2 \rangle_{\hbar} 
\langle (\delta \wh{p})^{2} \rangle_{\hbar} = \hbar^2/ 4$, 
which is reminiscent of Equation \eqref{eq:thermodynamic_uncertainty}. 
On the other hand, the linear correlation coefficient 
$\langle (\delta \wh{q} \delta \wh{p} + \delta \wh{p} \delta \wh{q})/ 2\rangle_{\hbar} 
\langle (\delta \wh{q})^2 \rangle_{\hbar}^{-1/ 2}
\langle (\delta \wh{p})^2 \rangle_{\hbar}^{-1/ 2}$
between 
$\delta \wh{q}$ and $\delta \wh{p}$ is equal to zero and not to $- 1$.

Compare this to the following. 
In classical mechanics, we have a concept of an ``action as a function of coordinates''. 
To simplify the discussion, take a one-dimensional harmonic oscillator 
with the Hamiltonian $H (p, q) = (p^2 + q^2)/2$, 
and look at the isoenergetic surface $L = \lbrace (p, q) \in \mathbb{R}^{2} \,|\, H (p, q) = c \rbrace$, 
where $c > 0$ is a parameter. 
$L$ is a one-dimensional Lagrangian manifold with respect to $d p \wedge dq$, 
and in a $q$-chart $U \subset L$, it can be described as 
$U = \lbrace (p, q) \,|\, p = \partial S (q)/ \partial q \rbrace$, 
where $S (q)$ is the action as a function of $q$, $d S = p d q$ on $L$. 
If $\alpha \in L$ is a point representing the state of the system at time $t = 0$, 
then after a short interval of time $\Delta t$, the system moves from 
$q (\alpha)$ to $q (\alpha) + \Delta q$, but one may say 
that the momentum, $p$, ``immediately adjusts'' its value 
from $p (\alpha)$ to $p (\alpha) + \Delta p$, 
so that the point stays on $L$. 
In the linear approximation: $\Delta p \simeq S'' (q (\alpha)) \Delta q$. 
This equation on its own does not prevent us from 
studying 
the quantum mechanics of our system. 
One may construct, for example, a semiclassical wavefunction of the shape: 
\begin{equation}
\label{eq:semiclassical_wavefunction}
\Psi_{\hbar} (x) = \int_{L} d \sigma (\alpha) \varphi (\alpha) 
\psi_{\hbar} (x; p (\alpha), q (\alpha), 1) 
\end{equation}
where $d \sigma$ is the measure on $L$ induced by $dp dq$ 
and $\varphi$ is a \emph{complex} smooth function 
on $L$ with a finite support. 
In short, $\Delta p \simeq S'' (q (\alpha)) \Delta q$ is a \emph{classical} equation, and, similarly, 
$\Delta \beta \simeq (\partial^{2} S/ \partial E^{2})_{\nu} \Delta E$ corresponds to 
the physics \emph{after} the thermodynamic limit. 

We can see that an analogy between Equation \eqref{eq:thermodynamic_uncertainty} and the Heisenberg uncertainty relation 
is not \emph{a priori} excluded, but it is far from being straightforward. 
Let us therefore clarify the concept of intensive quantities in thermodynamics 
as it is accepted in the present paper by considering the definition of inverse absolute temperature. 
I assume that one is well aware of what is a canonical Gibbs distribution. 
Naively, in the standard notation, the inverse absolute temperature is the parameter, 
$\beta$, in this distribution. 
In the opinion of the author, 
this is \emph{not} a conceptually correct way to define temperature. 
My starting point of view on thermodynamics is expressed in the book of M. Planck 
``Treatise on Thermodynamics''~\cite{Planck}. 
One \emph{should} (not just \emph{can}) 
start with phenomenological thermodynamics 
and perceive temperature (along with other quantities, like pressure, volume, \textit{etc}.) 
as an independently defined phenomenological concept. 
It is wrong to derive the concept of temperature from a statistical model (the Gibbs distribution). 
The Gibbs distribution is just a possible explanation and an 
interpretation of what we see in terms of an underlying multiparticle mechanical system. 
This is captured by the term 
``mechanical theory of heat'' which was more common in the old days than now.

Similarly, with the fluctuations, $\delta \beta$, 
the \emph{phenomenology} is described by the Einstein formula in 
Equation \eqref{eq:fluct_delta_beta}, 
but the ``mechanical theory'' of such fluctuations is, in my opinion, quite fuzzy. 
The aim of the present paper is to change this state of affairs a little. 
It is worth mentioning that in~\cite{Mandelbrot}, 
B.~Mandelbrot derives a kind of uncertainty relation for statistical estimators of thermodynamic quantities. 
I do not go in this direction in the present paper, 
but I consider essentially the same problem: 
what is the mechanical theory beyond the Einstein formula for $\delta \beta$?

The only reasonable possibility that I see 
to get some insight about such a theory is to use \emph{symmetry}. 
The Equations 
\eqref{eq:fluct_delta_varepsilon} and 
\eqref{eq:fluct_delta_beta}, 
for the fluctuations, $\delta \beta$ and $\delta \varepsilon$, 
are completely similar. 
For the fluctuations, $\delta \varepsilon$, we have a ``better'' theory 
given by the canonical Gibbs distribution at inverse absolute temperature $\beta (\alpha)$, $\alpha \in \Lambda$, and 
for the fluctuations, $\delta \beta$, we do \emph{not} have one. 
It is natural to try to restore this symmetry in a ``better'' theory. 
One may try to construct an auxiliary thermodynamic system 
similar to the one that we have and define the notation, 
$\Lambda'$, $\varepsilon'$, $\beta'$, $N'$, 
in a totally similar way to 
$\Lambda$, $\varepsilon$, $\beta$, $N$. 
The idea is as follows: for a fixed point, $\alpha \in \Lambda$, 
there is a point, $\alpha' \in \Lambda'$, such that:
\begin{equation} 
\langle 
(N^{1/ 2} \delta \beta)^{n}
\rangle_{\text{{\it ``better theory''}}, N, \alpha} = 
\langle 
((N')^{1/ 2} \delta \varepsilon')^{n}
\rangle
\end{equation}
for $n = 2, 3, 4, \dots$, 
where $\langle - \rangle$ 
on the right-hand side 
corresponds to the 
\par\noindent
averages computed using the canonical Gibbs distribution for $N' \in \mathbb{Z}$ particles at 
the inverse temperature $\beta' = \beta' (\alpha')$.

\section{Duality of Fluctuations}

As implied by 
the previous section, 
the idea of a ``quantum complementarity'' for the fluctuations of 
thermodynamic quantities should be handled with care. 
For a recent discussion, see \cite{Velazquez, VelazquezCurilef, UffinkLith}. 
Let us analyze this problem 
in terms of 
the inverse absolute temperature, $\beta$, and specific internal energy $\varepsilon$. 
We keep the notation for the Lagrangian manifold: 
$\Lambda \subset \mathbb{R}^{2} (\beta, \varepsilon)$. 
Consider a pair of cases: 
\begin{itemize}
\item[(i)] The system is in a thermostat. 

\item[(ii)] The system is adiabatically isolated from the environment. 

\end{itemize}

Case (i) is describing a parameter, $\beta_{*}$, which is 
the inverse absolute temperature of the thermostat. 
For an $N$-particle system, 
we have a Gibbs distribution: 
\begin{equation} 
\label{eq:Gibbs_dist}
w_{n}^{(N)} (\beta_*) = \frac{1}{Z_{N} (\beta_{*})} \exp (- \beta_{*} u_n^{(N)}/ k_{B}) 
\end{equation}
where 
$n$ is labeling all possible quantum states of the system, 
$w_n^{(N)} (\beta_*)$ is the probability of finding the system in a state, $n$, 
$u_n^{(N)}$ is the energy of the system in this state 
and 
$Z_{N} (\beta_{*})$ is the partition function. 
Assume that the spectrum is discrete, 
$n = 0, 1, 2, \dots$, and that 
$\min_{n} u_n^{(N)} \geqslant 0$). 
If we look at the system from a \emph{phenomenological} level, then 
we say that the inverse absolute temperature of the system $\beta = \beta_{*}$. 
If we look from the level of \emph{statistical} thermodynamics, then 
we perceive the energy present in the system as a random 
variable; denote it 
$\mathcal{E}_{N, \beta_*}$. 
The possible values of 
$\mathcal{E}_{N, \beta_*}$ are 
$\lbrace u_{n}^{(N)} \rbrace_{n}$, and the corresponding probability weights are 
$\lbrace w_{n}^{(N)} (\beta_*) \rbrace_{n}$. 
The quantity, $\mathcal{E}_{N, \beta_*}$, fluctuates around the 
value 
$\langle \mathcal{E}_{N, \beta_*} \rangle = - k_B (\partial/ \partial \beta) \log Z_{N} (\beta) |_{\beta = \beta_{*}}$, 
where $\langle - \rangle$ denotes the mathematical expectation, 
with a nontrivial variance 
$\mathrm{Var} (\mathcal{E}_{N, \beta_*}) = 
(- k_B \partial/ \partial \beta)^{2} \log Z_{N} (\beta)|_{\beta = \beta_{*}}$. 
In the quasithermodynamic approximation, we have: 
\begin{equation} 
\langle \mathcal{E}_{N, \beta_{*}} \rangle = O (N), \quad 
\mathrm{Var} (\mathcal{E}_{N, \beta_{*}}) = O (N) 
\end{equation}
where $N \to + \infty$. 
In the notation of the Equation \eqref{eq:fluct_delta_varepsilon}, 
the point, 
$\alpha \in \Lambda$, is determined by $\beta (\alpha) = \beta_*$, 
$\varepsilon (\alpha) = \lim_{N \to + \infty} \langle \mathcal{E}_{N, \beta_{*}} \rangle/ N$,
the fluctuation, $\delta \varepsilon$ corresponds, to 
$(\mathcal{E}_{N, \beta_{*}} - 
\langle \mathcal{E}_{N, \beta_{*}} \rangle)/ N$, 
$N \in \mathbb{Z}$ and:
\begin{equation} 
s (\varepsilon) = \lim_{N \to + \infty} N^{-1} S^{(\mathit{stat})} (N, N \varepsilon) 
\end{equation}
where $S^{(\mathit{stat})} = S^{(\mathit{stat})} (N, E)$ is the Legendre transform of the function 
$\Phi^{(\mathit{stat})} (N, \beta) := k_B \log Z_{N} (\beta)$ 
in the variable, $\beta$: 
\begin{equation} 
S^{(\mathit{stat})} (N, E) = (\Phi^{(\mathit{stat})} (N, \beta) + \beta E)|_{\beta = \beta(N, E)} 
\end{equation}
where $\beta = \beta (N, E)$ is the solution of the equation 
$E = \partial \Phi^{(\mathit{stat})} (N, \beta)/ \partial \beta$ with respect to $\beta$. 
It is assumed that the corresponding limits and the Legendre transform exist.

Case (ii) is intuitively ``dual'' to case (i). 
The parameter, $\beta_{*}$, is replaced by 
the internal energy of the system, 
$E_{*}$. 
It is quite important to stress that $E_{*}$ 
is defined at the level of \emph{phenomenological} 
thermodynamics. In this sense, it has nothing to do with the quantum mechanical energy 
spectrum of the system, $\lbrace u_{n}^{(N)} \rbrace_{n}$, which is used in \emph{statistical} 
thermodynamics. The parameter, $E_{*}$, is \emph{not} a ``selected'' energy level, 
$u_{n_0}^{(N)} \in \lbrace u_{n}^{(N)} \rbrace_{n}$. 

It is a standard practice in the textbooks to replace Equation \eqref{eq:Gibbs_dist} with 
the \emph{microcanonical} Gibbs~distribution: 
\begin{equation} 
\label{eq:microcan_dist}
W_{n}^{(N)} (E_{*}; \delta) = \frac{1}{\Gamma_{N} (E_{*}; \delta)} \chi_{[- \delta/2, \delta/ 2]} (u_{n}^{(N)} - E_{*}) 
\end{equation}
where $W_{n}^{(N)} (E_{*}; \delta)$ are the probabilities of finding the system in the states, 
$n = 0, 1, 2, \dots$, 
$\delta > 0$ is a small parameter, 
$\chi_{[- \delta/2, \delta/ 2]}$ is the indicator function of the segment, $[- \delta/2, \delta/ 2]$, 
and $\Gamma_{N} (E_{*}; \delta)$ is the normalization constant termed the \emph{statistical weight}. 
The problem with Equation \eqref{eq:microcan_dist} is that it contains an arbitrary parameter $\delta > 0$, 
and one first needs to compute the asymptotic behavior of the system as $N \to + \infty$ in the thermodynamic limit and then 
take the limit, $\delta \to 0$. 

The quasithermodynamic Equations \eqref{eq:fluct_delta_varepsilon} and 
\eqref{eq:fluct_delta_beta} for the fluctuations, $\delta \varepsilon$ and $\delta \beta$, 
look very similar, while Equations \eqref{eq:Gibbs_dist} and \eqref{eq:microcan_dist} are completely different. 
It is natural, on the other hand, to expect that the symmetry between $\delta \varepsilon$ and $\delta \beta$ 
stems from a deeper level of \emph{statistical} thermodynamics. 
Unfortunately, essentially, there is no theory of 
fluctuations of intensive thermodynamic parameters beyond the quasithermodynamic theory. 

It would be nice to have something like 
$\wt{w}_{m} \sim \exp (- E_{*} b_m/ k_B)$ for the probability, $\wt{w}_{m}$, of 
the inverse temperature in the system to have a value of $b_m$, $m = 0, 1, 2, \dots$. 
There is \emph{no} distribution like this that is known for a generic 
thermodynamic system, but, 
at the same time, nobody has proven that it cannot~exist. 

It is nonetheless completely legal to state the problem as follows. 
We may assume, without loss of generality, 
that 
$\beta$ and $\varepsilon$ 
have the same physical units: 
\begin{equation}
[\beta] = [\varepsilon] = [k_B^{1/ 2}] 
\end{equation}

Since $\beta$ is initially the inverse absolute temperature, we need to 
redefine it by multiplying by a constant factor, and 
similarly, we need to redefine the energy by dividing it by the same factor. 
The canonical Gibbs distribution in Equation \eqref{eq:Gibbs_dist} keeps its original shape. 
Imagine that we have another thermodynamic system of $N'$ particles 
with an energy spectrum, $\lbrace {u'}_{n}^{(N')} \rbrace_n$, $n = 0, 1, 2, \dots$, 
and the rescaled inverse absolute temperature, $\beta'$, such that $[\beta'] = [k_{B}^{1/ 2}]$. 
Define ${\mathcal{E}}'_{N', \beta'}$ in analogy with 
$\mathcal{E}_{N, \beta}$, 
replacing $\beta$ with $\beta'$ and 
$\lbrace u_{n}^{(N)} \rbrace_{n}$ with 
$\lbrace {u'}_{n}^{(N')} \rbrace_{n}$, $n = 0, 1, 2, \dots$. 
Let $\Lambda' \subset \mathbb{R} (\beta', \varepsilon')$ be the Lagrangian manifold for the second system 
defined similar to $\Lambda \subset \mathbb{R}^2 (\beta, \varepsilon)$. 
We state the problem as follows: 
adjust the state, $\alpha' \in \Lambda'$, and 
the parameters of the construction of the second system in such a 
way that the fluctuations of 
$\delta \varepsilon' := 
({\mathcal{E}}'_{N', \beta' (\alpha')} - \langle {\mathcal{E}}'_{N', \beta' (\alpha')} \rangle)/ N'$ 
approximate the fluctuations of the inverse temperature 
in the first system as well as possible. 
If $\alpha \in \Lambda$ is a selected point 
on the Lagrangian manifold, $\Lambda \subset \mathbb{R}^{2} (\beta, \varepsilon)$, 
of the first system, then we are interested in the fluctuations, $\delta \beta$, 
around the value $\beta = \beta (\alpha)$, and the ultimate goal is: 
\begin{equation} 
\label{eq:ultimate_goal}
\langle (\delta \beta)^{n} \rangle = 
\langle (
\delta \varepsilon'
)^{n} \rangle 
\end{equation}
for $n = 2, 3, 4, \dots$, where the 
moments of $\delta \beta$ on 
the left-hand side are determined by \emph{experiment}. 
If we restrict the range of possible values of $n$ as $n = 2, 3, \dots, n_0$, then, 
assuming the auxiliary thermodynamic system contains many enough parameters, 
the tuning Equation \eqref{eq:ultimate_goal} becomes possible. 

Before working out some examples, 
let us 
comment on 
the notation, $k_B \to 0$, 
mentioned in the Introduction. 
Suppose we have an entropy of the system, 
$S^{(\mathit{stat})} (E_0, E_1, \dots, E_d)$, 
in terms of the extensive coordinates $E = (E_0, E_1, \dots, E_d)$, 
which we have computed from the canonical partition function. 
Speaking of a thermodynamic limit, we are interested in the 
asymptotic behavior of a function: 
\begin{equation}
S_{\lambda}^{(\mathit{stat})} (E) := 
\lambda^{-1} S^{(\mathit{stat})} (\lambda E_0, \lambda E_1, \dots, \lambda E_d) 
\end{equation}
where $\lambda \to + \infty$ is a dimensionless large parameter. 
The limit $S (E) = \lim_{\lambda \to + \infty} S_{\lambda}^{(\mathit{stat})} (E)$ 
is the entropy in the phenomenological thermodynamics, and its gradient 
determines the Lagrangian manifold, $\Lambda^{0} \subset \mathbb{R}^{2 (d + 1)} (\wt{\beta}, E)$, 
$\wt{\beta} = (\beta_0, \beta_1, \dots, \beta_d)$. 
The scheme of the computations can be organized as follows: 
\begin{itemize}
\item[(1)] 
Put formally, $k_B = 1$ (this trick is similar to $\hbar = c = 1$). 

\item[(2)] 
Compute the required number of terms of the asymptotic expansion of 
$S_{\lambda}^{(\mathit{stat})} (E)$
as $\lambda \to \infty$. 

\item[(3)] 
Specialize $\lambda = 1$. 

\item[(4)] 
Recover the Boltzmann constant, $k_{B}$, given by Equation \eqref{eq:Boltzmann_constant} 
from the dimension analysis. 

\end{itemize}

Since at Step (2), the notation, $k_B$, is not reserved, we may re-denote 
$\lambda = (k_{B})^{-1}$ and speak of $k_B \to 0$. 
At Step (3), we then specialize $k_B = 1$, making the symbol, $k_B$, unreserved again.

\vspace{0.25 true cm}
\noindent

\emph{Example 1.}
Take a system of $N$ quantum harmonic oscillators with frequency $\omega$. 
Put $k_B = 1$. 
If we shift the ground energy level of an oscillator to zero, 
then the partition function of a single oscillator at inverse temperature $\beta$ 
is of the shape: 
\begin{equation} 
Z_{1} (\beta) = \sum_{n = 0}^{\infty} \exp (- \beta \bar \varepsilon_n) = 
(1 - \exp (- \beta a))^{-1} 
\end{equation}
where $a = \hbar \omega$, 
$\bar \varepsilon_n = a n$. 
Note that the parameters, 
$\beta$ and $a$, enter the formula $Z_1 (\beta) = \sum_{n = 0}^{\infty} \exp (- \beta a n)$ 
in the same way, and 
one can perceive it as a sum $\sum_{n = 0}^{\infty} \exp (- a \bar \beta_n)$, $\bar \beta_n = \beta n$. 
The partition function of $N$ oscillators is $Z_{N} (\beta) = (Z_{1} (\beta))^{N}$, 
and the corresponding derivatives yield: 
\begin{gather} 
\langle \mathcal{E}_{N, \beta} \rangle = 
N a (\exp (\beta a) - 1)^{-1} \\ 
\mathrm{Var} (\mathcal{E}_{N, \beta}) = 
N^{-1} \exp (\beta a) \langle \mathcal{E}_{N, \beta} \rangle^{2} 
\end{gather}

The entropy function 
$S^{(\mathit{stat})} = S^{(\mathit{stat})} (E_0, E_1)$, where 
$E_0 = N$ and 
$E_1 = E$ is the internal energy, is of the shape: 
\begin{equation} 
S^{(\mathit{stat})} (N, E) = N \Big\lbrace 
- \log \frac{a N}{E}
+ \Big( 1 + \frac{E}{a N} \Big)
\log \Big(1 + \frac{a N}{E} \Big)
\Big\rbrace 
\end{equation}

The Lagrangian manifold, $\Lambda \subset \mathbb{R}^{2} (\beta, \varepsilon)$, 
corresponding to the phenomenological specific entropy 
$s (\varepsilon) = \lim_{N \to + \infty} N^{-1} S^{(\mathit{stat})} (N, N \varepsilon)$ 
is described by an equation: 
\begin{equation} 
\beta = a^{-1} \log (1 + a/ \varepsilon) 
\end{equation} 

The Boltzmann constant is recovered as:
$\beta = k_{B} a^{-1} \log (1 + a/ \varepsilon)$ 
\mydiamond

\begin{prop}
Let $X$ be a system of $N$ quantum harmonic oscillators of frequency $\omega$ 
put in a thermostat at inverse absolute temperature $\beta$. 
Then, there exists a system, $X'$, of $N'$ quantum harmonic oscillators of 
frequency $\omega'$ and a value, $\beta'$, of the inverse absolute temperature in $X'$, 
such that the quasithermodynamic fluctuations, $\delta \beta$, in $X$ are described 
by the same probability density as the quasithermodynamic fluctuations of the 
specific internal energy, $\delta \varepsilon'$, in $X'$.
\end{prop}

\noindent 
\emph{Proof.} 
Put $k_B = 1$. Denote $a = \hbar \omega$, $a' = \hbar \omega'$. 
We need to satisfy a condition: 
\begin{equation} 
\mathrm{Var} (\mathcal{E}_{N, \beta}/ N) 
\mathrm{Var} ({\mathcal{E}}'_{N', \beta'}/ N') = 1 
\end{equation}
We have the unknowns $a'$, $\beta'$, $N'$ and the parameters, $a$, $\beta$, $N$. 
Impose a condition: 
\begin{equation} 
\label{eq:imposed_cond}
(\langle \mathcal{E}_{N, \beta} \rangle/ N) 
(\langle {\mathcal{E}}'_{N', \beta'} \rangle/ N') = \beta' \beta 
\end{equation}
and put $N' = N$. 
Taking into account Example 1, we obtain a system of equations: 
\begin{gather}
\label{eq:duality_system_eq1}
\frac{a}{\exp (\beta a) - 1} 
\frac{a'}{\exp (\beta' a') - 1} = \beta' \beta \\
\label{eq:duality_system_eq2}
(\beta' \beta)^{2} \exp (\beta a) \exp (\beta' a') = 1 
\end{gather}

Using the second equation, we derive from the first equation: 
\begin{equation} 
\frac{\beta' a'}{1 - \exp (- \beta' a')} = 
\frac{1 - \exp (- \beta a)}{\beta a}
\end{equation} 

The function $\varphi (z) = z/ (1 - \exp (- z))$, $z \in \mathbb{R}$ 
is a continuous and monotonic function, $\varphi (0) = 1$, such that 
$\varphi (z) \to 0$, if $z \to - \infty$, and 
$\varphi (z) \to + \infty$, if $z \to + \infty$. 
Our equation is of the shape $\varphi (\beta' a') = 1/ \varphi (\beta a)$, 
so if $\beta a$ is known, then $\beta' a'$ is uniquely determined. 
Once we know $\beta$, $a$ and $\beta' a'$, we can 
compute $\beta'$ as $\beta' = \beta^{-1} \exp (- \beta a/ 2) 
\linebreak
\exp (- \beta' a' / 2)$. 
In other words, for any $\beta > 0$ and $a > 0$, the system Equations \eqref{eq:duality_system_eq1} and \eqref{eq:duality_system_eq2} 
have a unique solution $(\beta', a') = (\beta_{0}', a_{0}')$ with respect to $\beta' > 0$ and $a' > 0$. 
The required system, $X'$, is described by $a' = a_0'$ and $N' = N$, 
and the corresponding inverse temperature is $\beta' = \beta_0'$. 
\myqed

\vspace{0.25 true cm}
\noindent
\emph{Remark 1.} 
The construction of $(X', \beta')$ is not uniquely determined. 
The additional Equation \eqref{eq:imposed_cond} can be chosen differently. 
Let, for example, 
$\langle {\mathcal{E}}'_{N', \beta'} \rangle/ N' = \beta$. 
Keeping $N' = N$, one arrives at a system: 
\begin{gather}
\label{eq:imposed_cond_other_eq1}
a' (\exp (\beta' a') - 1)^{-1} = \beta, \\
\label{eq:imposed_cond_other_eq2}
\exp (\beta a + \beta' a') 
\Big( 
\frac{\beta a}{\exp (\beta a) - 1}
\Big)^2 = 1 
\end{gather}
It follows that 
$\beta' a' = 2 \log (\sinh (\beta a/ 2)/ (\beta a/ 2))$. 
Substituting this into the first equation, one computes~$a'$. 
The system Equations \eqref{eq:imposed_cond_other_eq1} and \eqref{eq:imposed_cond_other_eq2} have a unique solution 
with respect to $(a', \beta')$. 
On the other hand, it does not follow that 
$\langle \mathcal{E}_{N, \beta} \rangle/ N = \beta'$, 
so the option Equation \eqref{eq:imposed_cond} is more symmetric. 
\mydiamond

\vspace{0.25 true cm}
Let us say that $(X', \beta')$ satisfying Proposition 1 is \emph{quasithermodynamically dual} to $(X, \beta)$. 
The quasithermodynamic 
fluctuations of intensive and specific extensive thermodynamic 
quantities 
Equations~\eqref{eq:fluct_delta_varepsilon} and \eqref{eq:fluct_delta_beta}
switch their roles if we 
switch between $(X, \beta)$ and $(X', \beta')$. 

\section{The Pauli Problem}

Let $X = X_{N} (a)$ be the thermodynamic system, $N$, oscillators of frequency $\omega = a/ \hbar$ 
in a thermostat at inverse temperature $\beta$. 
Denote $\mathcal{E}_{N, \beta}^{(a)}$ as the random variable 
corresponding to the energy contained in $X = X_{N} (a)$ at inverse temperature $\beta$. 
We have already computed $\langle \mathcal{E}_{N, \beta}^{(a)} \rangle$ and 
$\mathrm{Var} (\mathcal{E}_{N, \beta}^{(a)})$, but 
it is not difficult to 
find the higher order cumulants, as well: 
\begin{equation}
K_n [\mathcal{E}_{N, \beta}^{(a)}] := 
\Big( - \frac{\partial}{\partial \beta} \Big)^{n} \log Z_{N} (\beta; a) 
\end{equation}
where $n = 1, 2, 3, \dots$, 
$Z_{N} (\beta; a)$ is the partition function of the system $X = X_{N} (a)$, and 
we work in the system of units $k_B = 1$. 
Let us remind that if $t \to 0$ is a small parameter, then, for any $n \geqslant 1$, it holds: 
\begin{equation} 
1 + \sum_{m = 1}^{n} \frac{t^m}{m!} \langle (\mathcal{E}_{N, \beta}^{(a)})^{m} \rangle = 
\exp \bigg(
\sum_{m = 1}^{n} \frac{t^m}{m!} K_{m} [\mathcal{E}_{N, \beta}^{(a)}] 
\bigg) + O (t^{n + 1})
\end{equation}
as long as the corresponding moments exist. 
Expanding the exponent into a power series, one may recompute the cumulants 
in terms of the moments, and \emph{vice versa}.

Since 
$K_{n} [\mathcal{E}_{N, \beta}^{(a)}] = 
N a^{n} (- \partial/ \partial x)^{n - 1} (e^{x} - 1)^{-1} |_{x = \beta a}$, we obtain: 
\begin{equation} 
\label{eq:cumulants_E}
K_{n} [\mathcal{E}_{N, \beta}^{(a)}] = 
N a^{n} \sum_{m = 1}^{n} 
\frac{c (n, m)}{(e^{\beta a} - 1)^{m}} 
\end{equation} 
where $c (n, 1) = 1$, $c (n, n) = (n - 1)!$ and the 
coefficients, $c (n, m)$, satisfy a recurrent equation: 
\begin{equation} 
c (n + 1, m) = m c (n, m) + (m - 1) c (n, m - 1) 
\end{equation}
where $2 \leqslant m \leqslant n - 1$, and $n = 1, 2, 3, \dots$. 
This equation emerges in number theory in connection with the Bernoulli polynomials. 
For every pair of positive integers, $m$ and $n$, look at the sum:
\begin{equation} 
S_{m} (n) = 1^{m} + 2^{m} + \dots + n^{m} 
\end{equation}
This sum is known to be a \emph{polynomial} in $n$: 
\begin{equation} 
S_{m} (n) = \frac{1}{m + 1} \sum_{k = 0}^{m} \binom{m + 1}{k} B_{k} n^{m + 1 - k} 
\end{equation}
where $B_{k}$ are the Bernoulli numbers, 
$x/ (e^{x} - 1) = \sum_{n = 0}^{\infty} B_n x^{n}/ n!$. 
Another way to write the sum $S_{m} (n)$ is as follows: 
\begin{equation} 
S_{m} (n) = \sum_{k = 0}^{n - 1} \binom{n}{k + 1} c (m, k) 
\end{equation}
where $c (m, k)$ are the same coefficients as in Equation \eqref{eq:cumulants_E}. 
It is possible to describe them explicitly: 
\begin{equation} 
c (n, m) = \frac{1}{m} \sum_{k = 0}^{m} (-1)^{m - k} \binom{m}{k} k^{n} 
\end{equation}
for all $m = 1, 2, \dots, n$, and $n = 1, 2, 3, \dots$. 

In Proposition 1, we have constructed another system 
$X'$ of $N' = N$ 
oscillators with frequency 
$\omega' = a'/ \hbar$ at inverse temperature $\beta'$. 
Consider the cumulants of the energy in this system 
${\mathcal{E}}_{N', \beta'}^{(a')}$ at inverse temperature $\beta'$: 
\begin{equation}
K_n [{\mathcal{E}}_{N', \beta'}^{(a')}] := 
\Big( - \frac{\partial}{\partial \beta'} \Big)^{n} 
\log Z_{N'} (\beta'; a') 
\end{equation}
where $n = 1, 2, 3, \dots$. 
We know that $(X', \beta')$ and $(X, \beta)$ are quasithermodynamically dual, {\it i.e.}: 
\begin{equation} 
K_2 [\mathcal{E}_{N', \beta'}^{(a')}]
K_2 [\mathcal{E}_{N, \beta}^{(a)}] = N' N 
\end{equation}
Let us imagine for a short while that 
$(X', \beta')$ and $(X, \beta)$
are not just quasithermodynamically, but ``completely'' thermodynamically dual, 
{\it i.e.}, 
for every $n \geqslant 2$, 
the $n$-th cumulant of 
of the fluctuation, 
$\delta \beta$, 
of the inverse temperature in $X$ 
coincides with 
$K_n [\mathcal{E}_{N', \beta'}^{(a')}/ N']$. 
Equivalently, this implies that we know \emph{all} moments 
$\langle (\delta \beta)^n \rangle$, $n = 2, 3, 4, \dots$. 
In addition to the knowledge of the moments, $\langle (\delta \varepsilon)^{n} \rangle$, $n \geqslant 2$, for 
the fluctuations 
of the specific internal energy: 
\begin{equation} 
\delta \varepsilon = (\mathcal{E}_{N, \beta}^{(a)} - \langle \mathcal{E}_{N, \beta}^{(a)} \rangle)/ N 
\end{equation}
can we unite these data in a \emph{single} mathematical object? 

In quantum mechanics, the role of such an object is played by the wavefunction or, more generally, 
by the Wigner quasiprobability function. 
We wish to find $R_{N} (x, y) \in L^1 (\mathbb{R}^{2} (x, y))$, 
satisfying 
$\int_{\mathbb{R}^{2}} dx dy 
\linebreak
R_{N} (x, y) = 1$, 
such that: 
\begin{equation} 
\label{eq:moments_R}
\langle (\delta \varepsilon)^{n} \rangle = \int_{\mathbb{R}^{2}} d x d y \, x^{n} R_{N} (x, y), \quad 
\langle (\delta \beta)^{n} \rangle = \int_{\mathbb{R}^{2}} d x d y \, y^{n} R_{N} (x, y) 
\end{equation}
for $n = 1, 2, 3, \dots$. 
We do \emph{not} require $R_{N} (x, y)$ to be non-negative, but 
all that matters is that 
the integrals 
$R_{N}^{(1)} (x) := \int_{\mathbb{R}} d y\, R_{N} (x, y)$ and 
$R_{N}^{(2)} (y) := \int_{\mathbb{R}} d x\, R_{N} (x, y)$ 
can be perceived as usual probability densities. 

It is natural to consider a \emph{truncated} version of the 
system Equation \eqref{eq:moments_R} by requiring that these equalities hold only for $n \leqslant n_0$, where 
$n_0$ is a positive integer, termed the \emph{degree of truncation}. 
Then, we can always construct probability densities 
$f_{\delta \varepsilon}^{(n_0)} (x)$, $x \in \mathbb{R}$ and 
$f_{\delta \beta}^{(n_0)} (y)$, $y \in \mathbb{R}$, which are non-negative smooth 
real functions, such that:
\begin{equation}
\int_{\mathbb{R}} d x\, x^n f_{\delta \varepsilon}^{(n_0)} (x) = 
\langle (\delta \varepsilon)^{n} \rangle, \quad 
\int_{\mathbb{R}} d y\, y^n f_{\delta \beta}^{(n_0)} (y) = 
\langle (\delta \beta)^{n} \rangle 
\end{equation}
where $n = 1, 2, \dots, n_0$. 
The element, $R_{N} (x, y) \in L^1 (\mathbb{R}^{2} (x, y))$, should satisfy:
\begin{equation} 
\label{eq:Pauli_thermo}
f_{\delta \varepsilon}^{(n_0)} (x) = \int_{\mathbb{R}} d y \, R_{N} (x, y), \quad 
f_{\delta \beta}^{(n_0)} (y) = \int_{\mathbb{R}} d x \, R_{N} (x, y) 
\end{equation}
and it is necessary to impose some additional conditions on $R_{N} (x, y)$ to make this problem non-trivial. 

\begin{prop} 
Assume that 
$R_{N} (x, y) \in L^1 (\mathbb{R}^{2} (x, y))$ 
satisfies the system Equation \eqref{eq:Pauli_thermo} and that 
it is a symbol of a Weyl 
pseudo-differential operator 
$\wh{R}_{N} = R_{N} (x, - \ii \varkappa_{N} \partial/ \partial x)$ on 
$L^{2} (\mathbb{R} (x))$, where $\varkappa_{N} > 0$ is a constant. 
If $\wh{R}_{N}$ is an orthogonal projector onto a one-dimensional subspace, 
then $\varkappa_{N} \sim 2 k_{B}/ N$, as $N \to \infty$. 
\end{prop}

\noindent
\emph{Proof.} 
The parameter, $\varkappa_{N}$, is an analogue of $\hbar$ in the Pauli problem 
in quantum mechanics. 
Let us remind, that 
in one-dimensional quantum mechanics, the standard 
Pauli problem is defined as a problem of a reconstruction 
of a wavefunction, $\psi_{\hbar} (z) \in L^2 (\mathbb{R} (z))$, 
from the knowledge of $|\psi_{\hbar} (z)|^{2}$ and 
$|\wt{\psi}_{\hbar} (z)|^{2}$, where $\wt{\psi}_{\hbar} (z)$, $y \in \mathbb{R}$, 
is the $\hbar$-Fourier transform of $\psi_{\hbar} (z)$. 
Since $\delta \beta$ and $\delta \varepsilon$ are fluctuations of thermodynamic quantities 
around a point, $\alpha \in \Lambda$, on the Lagrangian manifold, 
$\Lambda \subset \mathbb{R}^{2} (\beta, \varepsilon)$, describing the system, 
they are approximated as $N \to + \infty$ 
by the normal distributions, 
$\delta \varepsilon \sim \mathcal{N} (0, k_B N^{-1} [- s'' (\varepsilon (\alpha))]^{-1})$ and 
$\delta \beta \sim \mathcal{N} (0, k_B N^{-1} [- s'' (\varepsilon (\alpha))])$, 
where $s = s (\varepsilon)$ is the specific entropy as a function of the specific internal energy, $\varepsilon$. 

Take a one-dimensional quantum system with a coordinate 
$\wh{q} = z$ and the corresponding momentum $\wh{p} = - \ii \hbar \partial/ \partial z$, 
and consider a 
Pauli problem with the densities of distribution of 
the coordinate and momentum 
given by the normal distributions, 
$\mathcal{N} (0, \lambda \hbar/ 2)$ and $\mathcal{N} (0, \hbar/ (2 \lambda))$, 
respectively, where $\lambda > 0$ is a parameter. 
The solution, $\rho_{\hbar} (q, p; \lambda)$, 
is known to exist, and it is given by the Wigner quasiprobability function 
associated with the wavefunction of the coherent state Equation 
\eqref{eq:coherent_state} concentrated in $p_0 = 0$, $q_0 = 0$. 
The operator $\wh{\rho}_{\hbar} (\lambda) = \rho_{\hbar} (z, - \ii \hbar \partial/ \partial z; \lambda)$ 
is a one-dimensional orthogonal projector on $L^2 (\mathbb{R} (z))$. 

If we formally replace 
the parameter, $\lambda$, with $[- s'' (\varepsilon (\alpha))]^{-1}$, 
$q$ with the coordinate, $x$, corresponding to $\delta \varepsilon$, 
$p$ with the coordinate, $y$, corresponding to $\delta \beta$, and 
$\hbar$ with $2 k_{B}/ N$, then 
we obtain a solution of the Pauli problem, $R_{N}^{(\mathit{quasi})} (x, y)$, for the 
fluctuations 
described by Equations \eqref{eq:fluct_delta_varepsilon} and
\eqref{eq:fluct_delta_beta}. 
The Weyl 
pseudo-differential operator 
$\wh{R}_{N}^{(\mathit{quasi})} = R_{N}^{(\mathit{quasi})} (x, - \ii 2 k_B N^{-1} \partial/ \partial x)$ 
is a one-dimensional orthogonal projector on $L^2 (\mathbb{R} (x))$. 
Comparing this to 
$\wh{R}_{N} = R_{N} (x, - \ii \varkappa_{N} \partial/ \partial x)$, 
we conclude: $\varkappa_{N} \sim 2 k_B/ N$, as $N \to \infty$. 
\myqed

\vspace{0.25 true cm}
Proposition 2 implies that 
if we are interested in the Pauli problem in thermodynamics, 
then the parameter $\varkappa = 2 k_B/ N$ plays a similar role 
to the semiclassical parameter, $\hbar \to 0$, in quantum mechanics. 
We have a self-adjoint operator on $L^2 (\mathbb{R} (x))$ with a unit trace: 
\begin{equation} 
\wh{R}_{N} = R_{N} \Big(x, - \ii \frac{2 k_B}{N} \frac{\partial}{\partial x} \Big) 
\end{equation} 
where 
$x$ and 
$- \ii 2 k_B N^{-1} \partial/ \partial x$ 
are Weyl ordered. 
There is an additional condition on $R_{N} (x, y)$. 
In the quasithermodynamic limit, $N \to \infty$, the symbol, 
$R_{N} (x, y)$, must be concentrated in the point 
$(x, y) = (0, 0)$: 
\begin{equation} 
\label{eq:Gauss_beta_varepsilon}
R_{N} (x, y) \sim \frac{N}{2 \pi k_B} \exp 
\Big\lbrace 
\frac{N (x^2 s'' (\varepsilon (\alpha)) + y^2 [s'' (\varepsilon (\alpha)) ]^{-1})}{2 k_B} 
\Big\rbrace 
\end{equation}
where 
$\alpha \in \Lambda$, $s'' (\varepsilon (\alpha)) < 0$.

\vspace{0.25 true cm}
\noindent
\emph{Remark 2.} 
It might seem that there is an essential difference between the fluctuations of $\beta$ and $\varepsilon$ 
in statistical thermodynamics and the fluctuations of $p$ and $q$ in quantum mechanics. 
In quantum mechanics, we have a \emph{freedom} of choice of what to measure, $p$ or $q$, 
but at the same time, we cannot choose to measure both. 
If the variance of $p$ tends to zero, then the variance of $q$ tends to infinity, and \emph{vice versa}, 
in accordance with the Heisenberg uncertainty relation. 
If we put a thermodynamic system, $X$, in a thermostat with inverse absolute temperature $\beta_{*}$, then 
we ``know'' the inverse absolute temperature 
in $X$, 
but at the same time, the variance of the 
fluctuation of $\varepsilon$ is \emph{finite}. 

In quantum mechanics, we should distinguish between 
two different stages: a \emph{preparation} of an experiment, 
and an \emph{interpretation} of the outcome of an experiment. 
These stages are separated by an \emph{act of measurement}. 
A preparation is described in the language of \emph{classical} mechanics 
({\it i.e.}, the conditions of experiment, the choice of a measuring device, \textit{etc}.). 
An outcome is described in the language of \emph{quantum} mechanics 
({\it i.e.}, the spectrum of an observable, quantum numbers, \textit{etc}.). 

We should perceive thermodynamics in a similar way. 
When we put the system, $X$, in a thermostat, we make a choice of a measuring device, {\it i.e.}, 
we say that we are going to \emph{observe} the fluctuations of $\varepsilon$. 
We speak of $\beta_{*}$ in the language of \emph{phenomenological} thermodynamics, 
and we interpret the observations of 
the fluctuations of $\varepsilon$ in the language of \emph{statistical} thermodynamics. 

If we surround the system, $X$, with adiabatic walls, 
then we make a choice of another ``measuring device''. 
We say that we are going to \emph{observe} the fluctuations of $\beta$. 
The result of our preparation for the experiment is 
the internal energy, $E_{*}$, for which we use the language of \emph{phenomenological} thermodynamics. 
The fluctuations of $\beta$ are observed in terms of \emph{statistical} thermodynamics. 

We cannot observe the fluctuations of $\varepsilon$ and $\beta$ both at the same time, just like 
we cannot do it with $p$ and $q$ in quantum mechanics. 
From the perspective of statistical thermodynamics, 
the symbols, $\beta_{*}$ and $E_{*}$, are \emph{parameters} of distributions describing observable quantities, like the fluctuations of $\beta$ and~$\varepsilon$. 
They are not ``observables'' themselves. 
What we \emph{can} do is not measuring, but constructing statistical estimators for $\beta_{*}$ or $E_{*}$ 
working with finite samples of measurement outcomes. 

One can find a much more detailed analysis of complementarity in statistical physics 
from a similar perspective in \cite{Velazquez}. 
In the terminology of that paper, it is important to distinguish between the variables 
that parametrize the \emph{existence} of the system
(e.g., the mechanical macroscopic observables), 
and the conjugated variables that are relevant for \emph{dynamical} descriptions 
of a tendency of the system to approach a thermodynamic equilibrium. 
\mydiamond

\vspace{0.25 true cm}
\noindent
\emph{Remark 3.} 
In principle, in thermodynamics, the number of particles, $N$, is large. 
On the other hand, we can \emph{formally} specialize $N = 1$ in the final formulae, and this 
leaves us with a rather ``weird'' entity: a thermodynamic system consisting of \emph{just one} 
particle. At the same time, this particle is still a mechanical particle, so it 
has some mechanical coordinates, $q$, and momenta, $p$, described by quantum mechanics. 
It follows that we arrive at some kind of hybrid object: it is a \emph{quantum} particle 
equipped with additional degrees of freedom of a \emph{thermodynamic} nature. 
There is, for example, an additional degree of freedom, $\beta$, which is 
the inverse absolute ``temperature'' of the particle. 
Let us term this $N = 1$ thermodynamic system a \emph{thermoparticle}. 

In thermodynamics, when one performs an intellectual leap 
from the phenomenological thermodynamics to the Gibbs distribution, 
one splits the system into a huge number of quantum particles, 
$\nu R = k_{B} N$; $\nu$ is the number of moles of the chemical substance of the system. 
Intuitively, it might be better to split the phenomenological thermodynamic system 
not into purely mechanical particles, 
but into thermoparticles. 
This is a ``natural speculation'' that goes far beyond the scope of the 
present article. 
\mydiamond

\vspace{0.25 true cm}
Let us now briefly consider 
from the perspective of the Pauli problem 
a possible construction of the fluctuation theory 
for $\delta \varepsilon$, $\delta \beta$, beyond the Gaussian approximation in Equations \eqref{eq:fluct_delta_varepsilon}, 
\eqref{eq:fluct_delta_beta} and 
\eqref{eq:Gauss_beta_varepsilon}. 
Take a one-dimensional quantum mechanical system 
with a coordinate $\wh{q} = x$ and the corresponding momentum $\wh{p} = - \ii \hbar \partial/ \partial x$ 
acting on the Hilbert space, $L^{2} (\mathbb{R} (x))$. 
The usual Pauli problem deals with a reconstruction of a wavefunction, $\psi_{\hbar} (x) \in L^{2} (\mathbb{R} (x))$, 
from the knowledge of $|\psi_{\hbar} (x)|^{2}$ and $|\wt{\psi}_{\hbar} (p)|^{2}$, where: 
\begin{equation} 
\wt{\psi}_{\hbar} (p) := (2 \pi e^{\ii \pi/ 2} \hbar)^{- 1/ 2} \int_{\mathbb{R}} dx \, \exp (- \ii p x/ \hbar) \psi_{\hbar} (x) 
\end{equation}
is the $\hbar$-Fourier transform of $\psi_{\hbar} (x)$. 

Suppose now that the state of the system is not necessarily pure, but is described 
by a density matrix, $\rho_{\hbar} (x, x')$. 
The corresponding Wigner quasiprobability function, $\mathcal{W}_{\hbar} (p, q)$, is computed as follows: 
\begin{equation} 
\mathcal{W}_{\hbar} (p, q) = \int_{\mathbb{R}} dx \, \rho_{\hbar} (q + x/ 2, q - x/2) \exp (- \ii p x/ \hbar) 
\end{equation}
The normalization condition $(2 \pi \hbar)^{-1} \int_{\mathbb{R}^{2}} d p d q\, \mathcal{W}_{\hbar} (p, q) = 1$. 
If the state is $\psi_{\hbar} (x)$, then 
$\rho_{\hbar} (x, x') = \psi_{\hbar} (x) \ov{\psi}_{\hbar} (x')$, 
where the bar denotes the complex conjugation. 
For the coherent state Equation \eqref{eq:coherent_state}, we have: 
$\mathcal{W}_{\hbar} (p, q) = 2 \exp (- \lambda (q - q_{0})^{2}/ (2 \hbar)) 
\exp (- 2 (p - p_{0})^{2}/ (\lambda \hbar))$. 
The value of an integral, 
$(2 \pi \hbar)^{-1} \int_{\mathbb{R}^{2}} dq dp\, \mathcal{W}_{\hbar} (p, q) q^{m} p^{n}$, 
where $m, n \in \mathbb{Z}_{+}$, 
yields an average corresponding to the 
Weyl ordering of the expression, $\wh{q}^{m} \wh{p}^{n}$; for example: 
$(2 \pi \hbar)^{-1} \int_{\mathbb{R}^{2}} dq dp\, \mathcal{W}_{\hbar} (p, q) q p = 
\langle (\wh{q} \wh{p} + \wh{p} \wh{q})/ 2 \rangle_{\hbar}$. 
Note that $(2 \pi \hbar)^{-1} \mathcal{W}_{\hbar} (p, q)$, 
considered as a replacement of 
the joint probability distribution density for 
the coordinate and the canonically conjugate momentum in 
classical mechanics, 
need not be non-negative, but it is real. 
Look at the observables: 
\begin{equation}
\wh{Z} (\mu, \nu) = \mu \wh{q} + \nu \wh{p} 
\end{equation}
where $\mu$ and $\nu$ vary over $\mathbb{R}$. 
Denote $\mathcal{T} (z; \mu, \nu)$ as the density of the probability that $\wh{Z} (\mu, \nu)$ 
has a value in $(z, z + dz]$. 
There exists a \emph{tomographic reconstruction formula} 
\cite{Ibort_Manko_Marmoc_Simonic_Ventrigliac, Mancini_Manko_Tombesi}: 
\begin{equation} 
\label{eq:tomographic_reconstruction}
\mathcal{W}_{\hbar} (p, q) = 
\frac{\hbar}{2 \pi} \int_{\mathbb{R}^{3}} 
\mathcal{T} (z; \mu, \nu) \exp \lbrace - \ii (z - \mu q - \nu p) \rbrace \, dz d\mu d \nu 
\end{equation}
Note that it suffices to have the data only for the observables:
\begin{equation} 
\wh{Z} (\cos (\theta), \sin (\theta)) = \cos (\theta) \wh{q} + \sin (\theta) \wh{p} 
\end{equation}
where 
$\theta \in [0, \pi)$, 
since the following should hold: 
\begin{equation} 
\mathcal{T} (z; \lambda \mu, \lambda \nu) = |\lambda|^{-1} \mathcal{T} (\lambda^{-1} z; \mu, \nu) 
\end{equation}
for any $\lambda \not = 0$. 

In principle, to step outside the Gaussian approximation in 
Equations \eqref{eq:fluct_delta_varepsilon} and~\eqref{eq:fluct_delta_beta}, 
one can mimic the reconstruction formula in Equation \eqref{eq:tomographic_reconstruction}. 
The problem here is not in whether or not one is ready to perceive 
$(\delta \varepsilon, \delta \beta)$ in analogy with quantum mechanics, 
but in the fact that we do not really use 
$\delta \xi = \mu \delta \varepsilon + \nu \delta \beta$. 
 It is nonetheless possible to obtain the quantum mechanical type formulae in a consistent way as follows.

Write the variances, $\langle (\delta \varepsilon)^{2} \rangle_{N, \alpha}$ and 
$\langle (\delta \beta)^2 \rangle_{N, \alpha}$, corresponding to Equations 
\eqref{eq:fluct_delta_varepsilon} and 
\eqref{eq:fluct_delta_beta}, as 
follows \cite{Ruuge1, Ruuge2, Ruuge3}:
\begin{equation}
\langle (\delta \varepsilon)^{2} \rangle_{N, \alpha} = 
\int_{\mathbb{R}} dx \, x^2 |\varphi_{h} (x)|^2, \quad 
\langle (\delta \beta)^{2} \rangle_{N, \alpha} = 
\int_{\mathbb{R}} dy \, y^2 |\wt{\varphi}_{h} (y)|^2 
\end{equation}
where $h = 2 k_B/ N$, 
$\varphi_{h} (x) = (\pi h/ \lambda)^{1/ 4} \exp (- \lambda x^2/ (2 h))$, 
and $\wt{\varphi}_{h} (y)$ is the $h$-Fourier transform of $\varphi_{h} (x)$. 
The thermodynamic ``wavefunction'', $\varphi_{h} (x)$, is just a square 
root of $f_{\delta \varepsilon} (x; \alpha, N)$, 
$\lambda = \lambda (\alpha)$, $\alpha \in \Lambda$. 
Let us perceive the $h$-Fourier transform $\wt{\varphi}_{h} (y)$ as follows: 
\begin{equation} 
\wt{\varphi}_{h} (y) = \int_{\mathbb{R}} d x\, G (y, x, \pi/ 2) \varphi_{h} (x) 
\end{equation}
where $G (y, x, t)$, $t > 0$, is the solution of the Cauchy problem: 
\begin{gather}
\ii \hbar \frac{\partial G}{\partial t} = 
\frac{1}{2} \Big(- h^2 \frac{\partial^2}{\partial x^2} + x^2 \Big) G \\ 
G|_{t = 0} = \delta (y - x) 
\end{gather}
The function, $G (y, x, t)$, is known explicitly: 
\begin{equation} 
G (y, x, t) = (2 \pi e^{\ii \pi/ 2} h \sin t)^{- 1/ 2} 
\exp \Big\lbrace 
\frac{\ii}{h} \Big( 
\frac{\cot t}{2} (y^2 + x^2) - \frac{y x}{\sin t}
\Big)
\Big\rbrace 
\end{equation}
Take a little more generic function than $\varphi_{h} (x)$: 
\begin{equation} 
\varphi_{h} (x; \lambda, x_0, y_0) = (\pi h/ \lambda)^{1/ 4} \exp \Big\lbrace 
\frac{\ii}{h} 
\Big[
\frac{\ii \lambda}{2} (x - x_0)^2 + y_0 x
\Big] \Big\rbrace 
\end{equation}
where $x_0, y_0 \in \mathbb{R}$ and $\lambda > 0$ are parameters. 
It is straightforward to check, that for: 
\begin{equation} 
\varphi_{h}^{t} (y; \lambda, x_0, y_0) := \int_{\mathbb{R}} d x\, 
G (y, x, t) \varphi_{h} (x; \lambda, x_0, y_0) 
\end{equation}
we obtain: 
\begin{equation} 
|\varphi_{h}^{t} (y; \lambda, x_0, y_0)|^{2} = 
(\pi h/ \lambda_t)^{1/ 2}
\exp \lbrace 
- \lambda_t (y - c_t)^2/ h
\rbrace 
\end{equation}
where:
$c_t = x_0 \cos t + y_0 \sin t$, and 
$\lambda_t = (\lambda^{-1} \cos^2 t + \lambda \sin^2 t)^{-1}$. 
Hence: 
\begin{equation} 
\begin{gathered}
\int_{\mathbb{R}} dx \, x |\varphi_{h}^{t} (x; \lambda, x_{0}, y_{0})|^{2} = c_t = x_0 \cos t + y_0 \sin t \\ 
\int_{\mathbb{R}} d x \, (x - c_t)^2 |\varphi_{h}^{t} (x; \lambda, x_{0}, y_{0})|^{2} = 
k_B N^{-1} (\lambda^{-1} \cos^2 t + \lambda \sin^2 t) 
\end{gathered}
\end{equation}
where we have substituted $h = 2 k_B/ N$. 

Consider now, as in Example 1 and Proposition 1, 
a system $X = X_N (a)$ of $N$ oscillators of frequency $\omega = a/ \hbar$ at inverse temperature $\beta$. 
The average $\ov{\varepsilon}$ of the specific internal energy, 
$\mathcal{E}_{N, \beta}^{(a)}/ N$, and its variance, 
$\langle (\delta \varepsilon)^2 \rangle$, 
$\delta \varepsilon := 
(\mathcal{E}_{N, \beta}^{(a)} - \langle \mathcal{E}_{N, \beta}^{(a)} \rangle)/ N$, 
which we compute using the canonical Gibbs formalism, 
are of the shape: 
\begin{equation} 
\ov{\varepsilon} = a (\exp (\beta a/ k_B) - 1)^{-1} 
\quad 
\langle (\delta \varepsilon)^2 \rangle = N^{-1} \exp (\beta a/ k_B) (\ov{\varepsilon})^{2} 
\end{equation}
Take another system, $X'$, of $N' = N$ oscillators with 
frequency $\omega' = a'/ \hbar$ at inverse temperature $\beta'$. 
Adjust the combination of parameters $(a', \beta')$ in such a way that 
$\langle (\delta \varepsilon)^{2} \rangle \langle (\delta \varepsilon')^{2} \rangle = k_{B}^{2} N^{-2}$, 
where $\delta \varepsilon'$ is the fluctuation of specific internal energy in $X'$. 
Construct now a ``homotopy'' from $X$ to $X'$. 
Let $X_t$, $t \in \mathbb{R}$ be a system of $N$ oscillators of frequency 
$\omega_t = a_t/ \hbar$ at inverse temperature $\beta_t$. 
Determine the parameters, $a_t$ and $\beta_t$, from the requirement: 
\begin{equation} 
\begin{gathered}
\ov{\varepsilon}_t = 
\ov{\varepsilon} \cos t + 
\ov{\varepsilon}' \sin t \\ 
\langle (\delta \varepsilon_t)^{2} \rangle = 
\langle (\delta \varepsilon)^{2} \rangle \cos^2 t + 
\langle (\delta \varepsilon')^{2} \rangle \sin^2 t 
\end{gathered}
\end{equation}
where $\ov{\varepsilon}_t$ is the average specific internal energy in $(X_t, \beta_t)$. 
Substituting the expressions 
$\ov{\varepsilon}_t = a_t (\exp (\beta_t a_t/ k_B) - 1)^{-1}$ and 
$\langle (\delta \varepsilon_t)^2 \rangle = N^{-1} \exp (\beta_t a_t/ k_B) ( \ov{\varepsilon}_t )^{2}$, 
we note that $N$ cancels out and that $a_t$ and $\beta_t$ are uniquely determined, 
for every $t \in \mathbb{R}$ and:
\begin{equation} 
(a_t, \beta_t)|_{t = 0} = (a, \beta), \quad 
(a_t, \beta_t)|_{t = \pi/ 2} = (a', \beta') 
\end{equation}

The right-hand side of the Equation \eqref{eq:cumulants_E}, if 
we substitute $\beta_t$ in place of $\beta$, and 
$a_t$ in place of $a$, 
determines the higher order cumulants $K_{n} [\mathcal{E}_{N, \beta_t}^{(a_t)}]$, 
$n = 3, 4, 5, \dots$, 
of the energy, $\mathcal{E}_{N, \beta_t}^{(a_t)}$ in $(X_t, \beta_t)$. 
One can always recompute these cumulants 
into the moments $\langle (\delta \varepsilon_t)^n \rangle$, 
$n = 2, 3, 4, \dots$, 
if necessary, $\langle \delta \varepsilon_t \rangle = 0$. 
Assume that for some $n_0 \in \mathbb{Z}$, $n_0 \geqslant 3$, we have 
constructed a function, $\mathcal{T} (z; \mu, \nu)$, such that:
\begin{gather} 
\mathcal{T} (z; \lambda \cos t, \lambda \sin t) = 
|\lambda|^{-1} \mathcal{T} (\lambda^{-1} z; \cos t, \sin t) \\ 
\langle (\delta \varepsilon_{t})^{n} \rangle = \int_{\mathbb{R}} d z\, z^{n} \mathcal{T} (z; \cos t, \sin t) 
\end{gather}
where $\lambda \not = 0$, $t \in [0, \pi)$, $n = 1, 2, \dots, n_0$. 
Then, it remains to apply the tomographic reconstruction formula in Equation \eqref{eq:tomographic_reconstruction} 
replacing $\hbar$ with $2 k_B/ N$. 
This yields 
a joint quasiprobability function describing $(\delta \varepsilon, \delta \beta)$, which is 
more advanced than the right-hand side of Equation \eqref{eq:Gauss_beta_varepsilon}. 

One should stress that this function is obtained by a mathematical analogy, and it 
certainly 
requires experimental tests and a theoretical extension to more realistic systems. 
This can be a subject of future research.

\section{Conclusions}

In phenomenological thermodynamics, as well as in the 
theory of quasithermodynamic fluctuations, 
there exists a certain symmetry between intensive and extensive
quantities. 
At the same time, 
this symmetry is not immediately visible 
in statistical thermodynamics, since 
the Gibbs formalism deals only with 
the fluctuations of extensive quantities (energy, number of particles, \textit{etc}.). 
An interpretation of the fluctuations of intensive quantities is 
a rather controversial issue.

In quasithermodynamics, the probability densities 
for the fluctuations 
of specific extensive quantities (e.g., $\delta \varepsilon$, where $\varepsilon$ is specific internal energy) 
and the associated intensive quantities (e.g., $\delta \beta$, where $\beta$ is inverse absolute temperature) 
are completely similar. 
In particular, the corresponding variances satisfy 
the same asymptotic estimates, 
$\langle (\delta \varepsilon)^{2} \rangle = O (N^{-1})$ and 
$\langle (\delta \beta)^{2} \rangle = O (N^{-1})$, as 
the number of particles $N \to + \infty$. 
Mathematically, it is possible to perceive the quasithermodynamic 
fluctuations of intensive quantities in a system $X$, 
as quasithermodynamic fluctuations of specific extensive quantities in 
another system, $Y$.

In the present paper, it is suggested to extend this fact 
to \emph{statistical} thermodynamics in order to construct a fluctuation theory 
of \emph{intensive} quantities. 
This turns out to be possible 
for a model system of $N$ quantum harmonic oscillators of the same frequency 
if one takes into account 
an analogy between 
the transition from quantum to classical mechanics and the 
transition from statistical to phenomenological thermodynamics. 
In the main text, we ``span'' a generalized Pauli problem 
over the fluctuations, $\delta \beta$ and~$\delta \varepsilon$, 
and apply the \emph{tomographic reconstruction} formula. 
This yields 
a self-adjoint non-negative operator, $\wh{R}$, with a unit trace 
on $L^{2} (\mathbb{R} (x))$:
\begin{equation} 
\wh{R} = R_{N} \Big(x, - \ii \frac{2 k_B}{N} \frac{\partial}{\partial x} \Big) 
\end{equation}
which is similar to the density matrix operator in quantum mechanics
(we use the Weyl quantization). 
The symbol, $R_{N} (x, y)$, replaces the Wigner function, and 
it is concentrated in a point $(x, y) = (0, 0)$ in the quasithermodynamic limit, $N \to + \infty$. 
The combination, $2 k_B/ N$, plays the same role as 
the Planck constant, $\hbar$.

\end{document}